\documentclass[aps,prl,twocolumn,footinbib,superscriptaddress]{revtex4-1}
\usepackage[cp1251]{inputenc}
\usepackage{amsmath}
\usepackage{amssymb}
\usepackage{color,xcolor}
\usepackage{natbib}
\usepackage[colorlinks=true,bookmarks=false,citecolor=blue,urlcolor=blue]{hyperref} 
\usepackage{bm}
\usepackage{epsfig}
\usepackage{verbatim} 
\usepackage{morefloats}
\usepackage{graphicx}
\usepackage{bibentry}
\usepackage[mathscr]{euscript}
\usepackage{enumerate}

\usepackage{subfigure}

\renewcommand{\phi}{\varphi}

\usepackage{bm}

\renewcommand{\Re}{\mathrm{Re}}
\renewcommand{\Im}{\mathrm{Im}}

\newcommand{\g}[1]{\textsl{g}_{#1}}

\begin{document}

\renewcommand{\Im}{\mathop{\mathrm{Im}}\nolimits}
\renewcommand{\Re}{\mathop{\mathrm{Re}}\nolimits}

\def\dfrac{\displaystyle\frac}  
\newcommand{\eps}{\varepsilon}
\newcommand{\vphi}{\varphi}

\newcommand{\epsg}[1]{\varepsilon_{{\bf #1}}} 

\newcommand{\commentYuri}[1]{{\color{olive} {\bf Yuri: }\it #1}}
\newcommand{\commentDasha}[1]{{\color{violet} {\bf Dasha:} \it #1}}

\title{Topological edge states and gap solitons in the nonlinear Dirac model
}   

\author{\firstname{Daria A.} \surname{Smirnova}}
\affiliation{Nonlinear Physics Centre, Australian National University, Canberra ACT 2601, Australia}
\affiliation{Institute of Applied Physics, Russian Academy of Science, Nizhny Novgorod 603950, Russia}

\author{\firstname{Lev A.} \surname{Smirnov}}
\affiliation{Institute of Applied Physics, Russian Academy of Science, Nizhny Novgorod 603950, Russia}

\author{Daniel Leykam}
\affiliation{Center for Theoretical Physics of Complex Systems, Institute for Basic Science (IBS), Daejeon 34126, Republic of Korea}

\author{\firstname{Yuri S.} \surname{Kivshar}}
\affiliation{Nonlinear Physics Centre, Australian National University, Canberra ACT 2601, Australia}

\begin{abstract}
Topological photonics has emerged recently as a novel approach for realizing robust optical circuitry, and the study of nonlinear effects in topological photonics is expected to open the door for tunability of photonic structures with topological properties. Here, we study the topological edge states and topological gap solitons which reside in the same bandgaps described by the nonlinear Dirac model, in both one- and two-dimensions.  We reveal strong nonlinear interaction between those dissimilar topological modes manifested in the excitation of the topological edge states by scattered gap solitons. Nonlinear tunability  of localized states is explicated with exact analytical solutions for the two-component spinor wave function. Our studies are complemented by spatiotemporal numerical modeling of the wave transport in one- and two-dimensional topological systems.  
\end{abstract}

\maketitle

The grand vision of robust waveguiding and routing of light motivates studies of topological photonic systems~\cite{Lu2016,Ozawa2019}. Nonlinearities in photonics and related fields such as acoustics and Bose-Einstein condensates are being pursued to combine topological protection with advanced functionalities including active tunability, genuine non-reciprocity, frequency conversion, and entangled particle generation~\cite{Zhou2017,Kartashov2017, Segev2018b,Chen2018,Mittal2018,Kruk2018,Leykam2018,Wang2019}. However, rigorously-defined topological invariants are presently restricted to linearized perturbations to nonlinear steady states~\cite{Bardyn2016,Bleu2016,Peano2016} and there is a pressing need to better understand nonlinear topological systems in non-perturbative regimes. 

Non-perturbative studies of nonlinear edge states and solitons have largely focused on the dynamics along the edge and been limited to numerical simulations, due to the scarceness of exact solutions to nonlinear problems~\cite{Segev2013,Leykam2016,Kartashov2017,Solnyshkov2017}. A notable exception are models based on nonlinear coupling, which can be understood semi-analytically in terms of nonlinearity-induced domain walls~\cite{Gerasimenko2016,Hadad2016,Hadad2017,Hadad2018,Zhou2017,Leykam2018,RingSoliton2018}. Although realized in electronic circuits~\cite{Hadad2018,Wang2019}, this class of models does not admit strictly localized nonlinear modes and is challenging to scale to optics, where local on-site Kerr nonlinearities are more feasible. Recent experiments using coupled optical fibre loops have shown that the latter class of nonlinearities can be used to couple between localized topological edge states and bulk modes~\cite{Bisianov2018}. So far, however, there were no systematic studies of this phenomenon and the nonlinear dynamics transverse to the edge. This paper is aimed at bridging this gap.

\begin{figure} [b!]
\centerline{\includegraphics*[width=\linewidth]{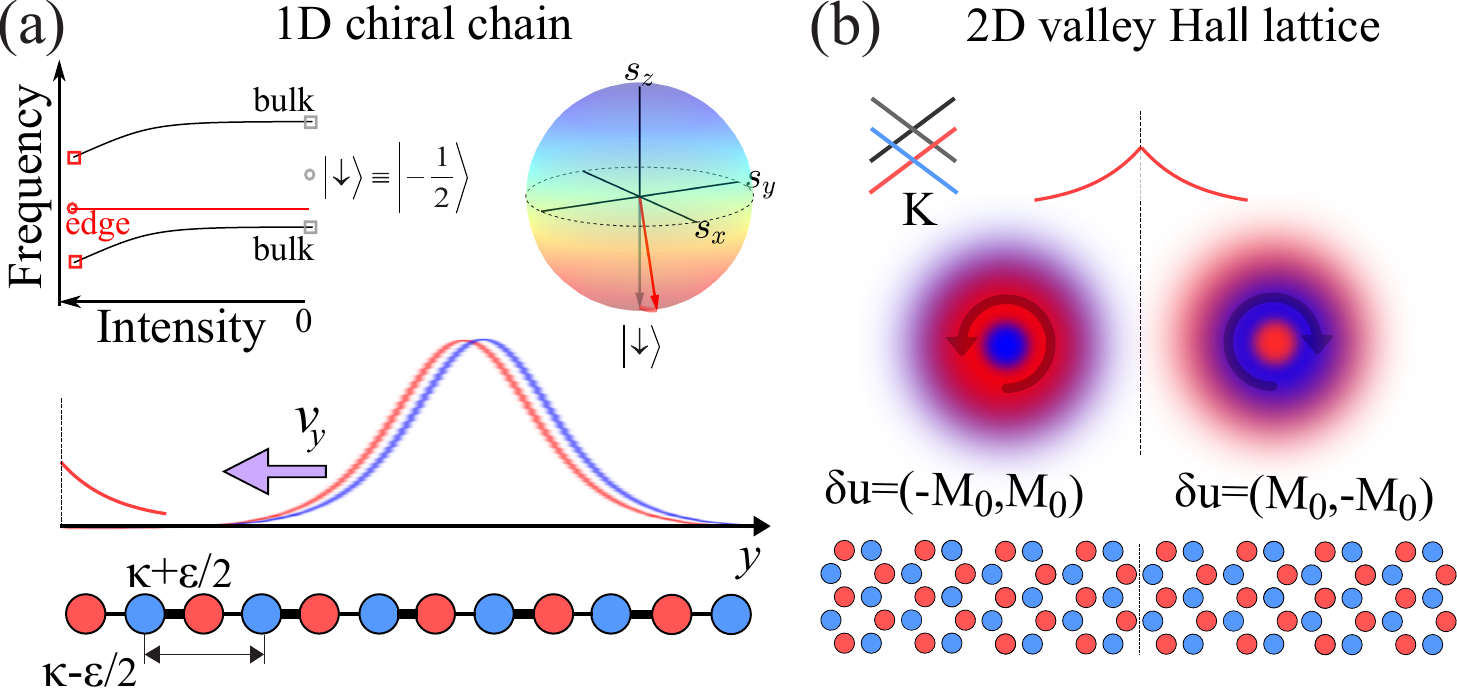}} 
\caption{Applications of the nonlinear Dirac model. 
(a) Gap soliton excites topological edge state of a 1D dimerized chain. Left inset: Schematic 
of the intensity-controlled tunability of the edge state frequency. Right inset: Bloch sphere representation of 
edge state spin polarization, tilted due to nonlinearity.
(b) Semi-vortex bulk solitons and edge states at a valley-Hall domain wall in graphene with staggered sublattice potential. 
} 
 \label{fig:fig1}
\end{figure}

Here we study a generic 
continuum Dirac model with local Kerr nonlinearity describing dimerized chains of optical resonators or photonic graphene with staggered sublattice potentials, illustrated in Fig.~\ref{fig:fig1}. We obtain 
localized nonlinear modes analytically, showing that bulk gap solitons and edge states have a unified origin as heteroclinic orbits of a nonlinear Hamiltonian system, differing only in their boundary conditions. In the nonlinear system, the mid-gap frequency appears a critical point of a 
phase transition where the topological edge states emerge.
Further numerical simulations reveal that scattering of travelling bulk solitons off the edge can efficiently excite topological edge states and there is an optimal soliton velocity maximizing the energy transfer. Our model is highly relevant to recent theoretical and experimental studies~\cite{Solnyshkov2017,Dobrykh2018}, with our exact solutions more informative and accurate than previous approximate variational approaches. This allows us explicate nonlinear tunability and transport properties of spin-polarized localized nonlinear modes residing in the nontrivial band gap.

We begin with the nonlinear Dirac model 
describing evolution of a spinor wavefunction $\Psi = [\Psi_{1},\Psi_{2}]$: 
\begin{align}
&i \partial_t \Psi = ({\hat H_D}(\delta \bm k) + {\hat H_{N\!L}}) \Psi, \label{eq:2D_Dirac_NL} \\
&{\hat H_D}(\delta \bm k)= \delta k_x \hat{\sigma}_x + \delta k_y \hat{\sigma}_y  + M \hat{\sigma}_z \:, \label{eq:H_D}
\end{align}
where $ \delta  {\bm k} = (\delta k_x, \delta k_y) \equiv 
- i (\partial_x, \partial_y ) $ is the momentum and
${\hat H_{N\!L}}= - g \bigl[{ |\Psi_{1}|^{2} }, {|\Psi_{2}|^{2} }\bigr]$ is a local cubic nonlinearity. We reduce  Eq.~\eqref{eq:2D_Dirac_NL}  to a quasi-1D form by considering plane wave-like states along the $x$ axis, 
$\Psi = \psi e^{ i k x}$, treating $k$ as a parameter,
\begin{equation} \label{eq:DiracEq1}
i\partial_t \psi \! = \left( \!{\begin{array}{*{22}{c}}
{M - g |\psi_1|^2} & {k -  \partial_y }\\
{k + \partial_y}  & {- M - g |\psi_2|^2}
\end{array}} \! \right) \psi.
\end{equation}
Conserved quantities are the power $\mathcal{P} = \langle{\bm{\psi}}|{\bm{\psi}}\rangle$ and the energy 
$\mathcal{E} = \langle{\bm{\psi}}|\left({\hat {H}_D + \frac{1}{2}\hat {H}_{N\!L}}\right)|{\bm{\psi}}\rangle$, where the inner products denote integration over $y$.

We seek stationary states with time dependence $\sim \exp(- i\omega t)$ and velocity $v$ by applying the Lorentz transformation,
\begin{equation}
\tau=\gamma\left(t-vy\right)\!,\: 
\xi=\gamma\left(y-vt\right)\!, \:  
\gamma=1\bigl/\sqrt{1-v^{2}}\bigr..
\end{equation}
Recasting the nonlinear eigenvalue problem as a Hamiltonian system governing the transverse mode profiles (Supplemental Material), we find that \emph{all} localized modes (edge or bulk) must have vanishing Hamiltonian. This leads to the closed form solution,
\begin{subequations} \label{eq:soliton_spinor}
\begin{gather}
\psi^{s}_{1}(\xi)\!=\!\sqrt{\frac{\rho_s}{2} } \left( e^{-i\alpha_s} + \gamma (1+v) e^{i\alpha_s} \right)e^{i\beta_s},\\
\psi^{s}_{2}(\xi)\!=\!- i \sqrt{\frac{\rho_s}{2}}  \left( e^{-i\alpha_s} - \gamma (1+v) e^{i\alpha_s} \right) e^{i\beta_s}, 
\end{gather}
\end{subequations} 
where $\rho_s$, $\alpha_s$, and $\beta_s$ are intensity, spin angle, and phase profiles, respectively, which take the form
\begin{eqnarray} 
& \alpha_{n}\left(\xi\right)=\tan^{ -1} \left[\frac{\Omega-\omega}{\sqrt{\Omega^{2}-\omega^{2}}}\tanh\left(\sqrt{\Omega^{2}-\omega^{2}}\,\xi\right)\right], \nonumber  \\
& \alpha_s(\xi) = \alpha_n  - \delta/2, \quad \varrho_{s} (\xi) = \dfrac{2(\Omega\cos{2\alpha_{n}}-\omega)}{{\mathcal B} + {\mathcal A}\cos^{2} {2\alpha_s }},  \nonumber \\
& \beta_{s}(\xi)=v\sqrt{\frac{\mathcal{B}}{\mathcal{A}+\mathcal{B}}}\tan^{ -1} \left(\sqrt{\frac{\mathcal{B}}{\mathcal{A}+\mathcal{B}}}\tan2\alpha_{s} \right),  
\end{eqnarray}
with coefficients 
$\mathcal{A}=\gamma\left(1+v\right)\g{}$,
$\mathcal{B}=\gamma\g{}\bigl/\left(1-v\right)$, 
$\Omega =  \sqrt{ k^2 + M_0^2 }$, 
$\delta = \tan ^{ - 1} {\left( \dfrac{k}{M_0}\right)}$, and we have taken $M = M_0 \ge 0$. The $v=0$ limit returns 
spinor components of the chiral soliton~\cite{Solnyshkov2017},
\begin{equation}
\left( \!{\begin{array}{*{20}{c}}
{ \psi^s_1 (y) } \\
{ \psi^s_2 (y)}
 \end{array}} \!  \right) =\sqrt{2\varrho_s (y)} e^{-i\omega (\mathcal{P}) t}\left( \!{\begin{array}{*{20}{c}}
{ \cos \alpha_s (y)} \\
{ -\sin \alpha_s (y)}
\end{array}} \!  \right),
\end{equation}
where the frequency $\omega (\mathcal{P})$ implicitly depends on the total 
power. Remarkably, this exact solution describes both bulk and edge solitons, with the latter interpreted as part of a stationary bulk soliton bound to the edge distinguished by the boundary conditions. 

\begin{figure} [t!]
\centerline{\includegraphics*[width=0.99\linewidth]{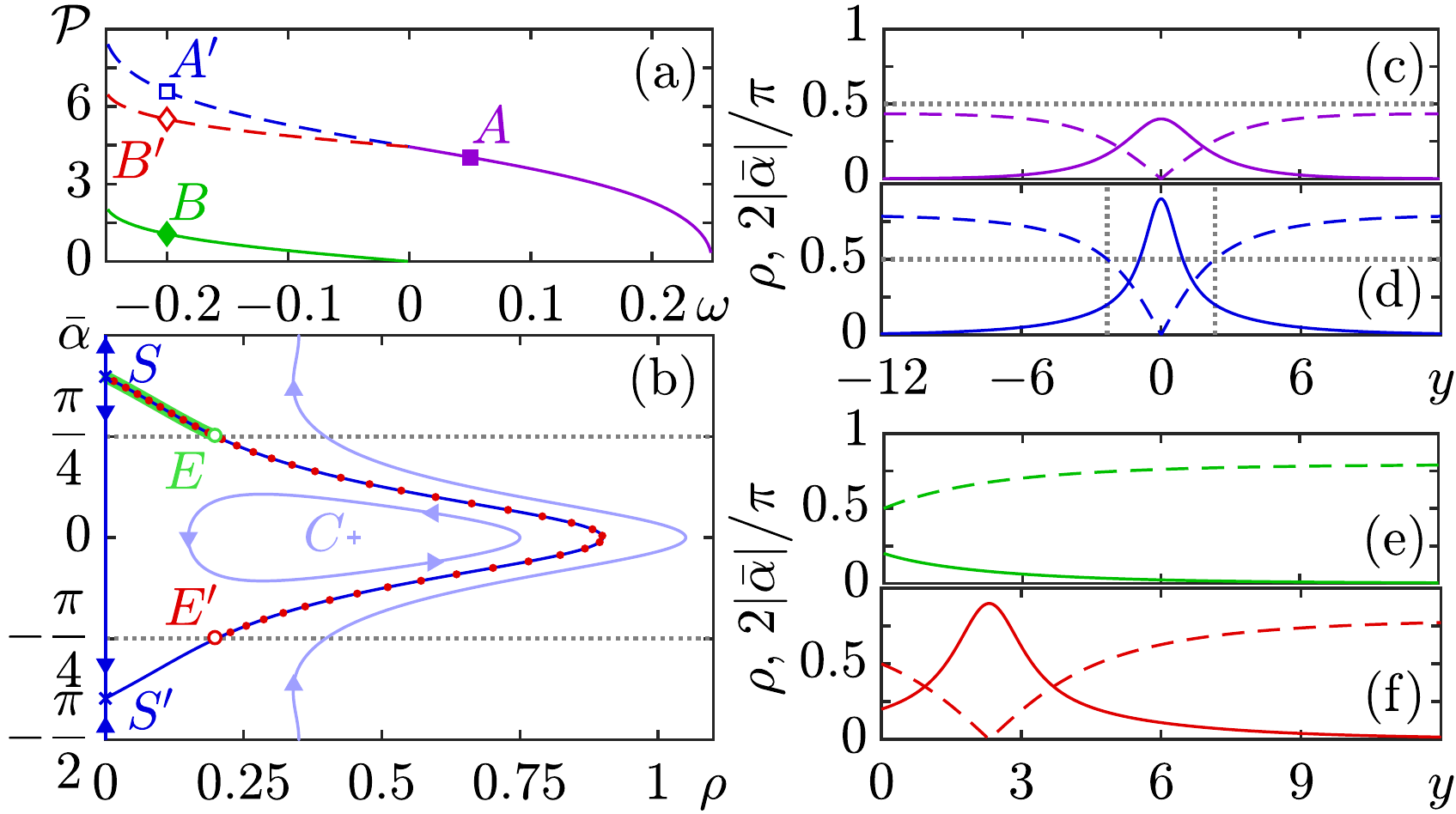}} 
\caption{ (a) Nonlinear dispersion of stationary localized states: dependence of the frequency on the total power
$\omega({\mathcal{P}})$ for bulk solitons (violet solid for stable and dashed blue for unstable branches) and families of the edge states (solid green for stable localized at the half-dimer and dashed red for unstable nonlinearly induced at the full dimer edge). (b) Phase portrait of the nonlinear SSH model, $v=0$, $\omega = - 0.075$ identify phase trajectories for soliton (blue) and edge states (green and red dotted).
(c,d) Stationary bulk solitons: profiles of (c) stable and (d) unstable solitons indicated by letters $A$ and $A'$ in panel (a). 
(e,f) Nonlinear edge states: profiles for (e) half-dimer ($B$) and (f) full-dimer ($B'$) 
chain termination at the left edge. (c-f) The intensity $\rho$ is plotted with solid lines, dashed lines correspond to $2|\bar{\alpha}|/\pi$. 
Parameters $\kappa=1$, $v_F = 1$, $g=1$, $\eps = 0.25$. 
}
 \label{fig:fig2}
\end{figure}
To illustrate the features of these exact solutions, we consider as an example the Su-Schrieffer-Heeger (SSH) model, describing a 1D dimer chain with alternating weak and strong couplings $\kappa_{1,2} = \kappa \pm \varepsilon/2$ between the nearest neighbors. Near the Brillouin zone edge $k_yd=\pi$ the bulk spectrum is 
captured by the continuum Dirac Hamiltonian 
\begin{equation} 
{\hat H_D} (\delta k)=  -\eps \hat{\sigma}_x + v_F \hat{\sigma}_y  \delta k, \label{SSH_spectrum}
\end{equation}
where $v_F = \kappa d $ is Fermi velocity and $d$ is the lattice period. Introducing self-focusing nonlinearity, this corresponds to Eq.~\eqref{eq:DiracEq1} with $k = -\eps$ and $M = 0$.

The phase plane ($\varrho, \bar{\alpha})$ of the Hamiltonian system determining the mode profiles is shown in Fig.~\ref{fig:fig2}(b), where $\bar{\alpha} = \alpha_s - \pi/4$. There are three equilibrium points: two saddles $S, S'$
$\rho=0$, 
 $\omega= \eps \sin(2\alpha)$, 
and center $C$
$\alpha = \pi/4$, $\rho = {(\eps - \omega)}/{g}$.
This phase portrait supports a bright soliton as a heteroclinic trajectory at vanishing Hamiltonian, corresponding to a separatrix between two saddles. In contrast to the nonlinear coupling model studied in Refs.~\cite{Hadad2016,Hadad2017,Hadad2018}, the bounded trajectory here is unique. 
The boundary conditions (i) $\alpha_s = \pi/2$ and (ii) $\alpha_s = 0$ describe linear edge states modified by nonlinearity and nonlinearity-induced edge states respectively. We verified these analytical solutions against stationary solutions found numerically using Newton's method. 

Integrating the intensity of the soliton solutions Eq.~\eqref{eq:soliton_spinor}, we obtain their total power and energy
{\footnotesize{
\begin{subequations}
\begin{align}
& {\mathcal P}_s (\omega, v) = \dfrac{4}{g \gamma \sqrt{1+\gamma^2}} \left[ \dfrac{\pi}{2} - \tan ^{ - 1} \left( {\dfrac{1}{\sqrt{1+\gamma^2}} \dfrac{\omega_s}{\sqrt{\eps^2 - \omega^2}}} \right) \right] \\
& \mathcal{E}_s (\omega, v) = \omega_s (1 - \gamma^2) {\mathcal P}_s + \dfrac{4 \gamma^2 \eps }{g} \tan ^{ - 1} \left( \dfrac{\sqrt{\eps^2 - \omega^2}} {\gamma \eps} \right)  \:,
\end{align}
\end{subequations}}} 
where $\omega_s = \gamma \omega$ is the soliton frequency in the laboratory frame.  In the linear limit $\omega \rightarrow \eps$ the energy of nonlinear interaction vanishes and $\mathcal{E}_s \rightarrow \omega_s {\mathcal P}_s $. We similarly obtain the type (i) edge state dispersion
\begin{equation}
{\mathcal P}^{(1)} (\omega) = \dfrac{\sqrt{2}}{g} \tan ^{ -1} {\dfrac{|\omega|}{\sqrt{2 (\eps^2 - \omega^2)}}},
\end{equation}
with asymptotic values ${\mathcal P}^{(1)} (\omega \rightarrow 0) \rightarrow 0 $ and ${\mathcal P}^{(1)} (\omega \rightarrow -\eps) = \pi / g \sqrt{2}$. The power of the type (ii) edge state 
can then be calculated as the difference ${\mathcal P}^{(2)} =  {\mathcal P}_s(\omega,v=0) -  {\mathcal P}^{(1)} $ and has a nonzero threshold
${\mathcal P}^{(2)} (\omega \rightarrow 0) \rightarrow \sqrt{2} \pi /g $. 
The modal dispersion plotted in Fig.~\ref{fig:fig2}(a) shows that the type (i) edge states bifurcate from the linear midgap topological edge modes at zero intensity, indicating the absence of any threshold necessary for their existence. They exist in the range $-\eps<\omega (\mathcal{P}) \le 0$ that can be controlled by dimerization strength $\eps$. Notably, $\omega = 0$ coincides with a bifurcation of the bulk soliton, which loses its stability and produces a type (ii) nonlinear edge state. 

To relate these observations to the topological properties of the SSH model it is useful to study the soliton intensity and spin angle profiles ($\rho,\alpha)$, shown in Figs.~\ref{fig:fig2}(c-f). The latter determines the $y$-dependent spin polarization density ${\bm S} = \frac{1}{2} {\bm{\psi}}^\dagger {\hat{{\bm \sigma}}}{\bm{\psi}}$. For solitons at rest this yields 
${\bm S}_s (y)= \rho_s (-\sin{2\alpha_s},0,\cos{2\alpha_s})$. The spin angle $\bar{\alpha}$ changes sign at the soliton core and localization requires $\bar{\alpha}$ to asymptotically approach $\cos 2\bar{\alpha} = \omega / \eps$. The bulk solitons hence have total spin $\langle {\bm{S}}  \rangle_s  = \int {\bm{S}} \,\mathrm{d}y = -{2}/{g} \left( \tan^{-1} \left( {\sqrt{1-{\omega^2}{\eps^{-2}}}}\right),0,0\right)$, which respects the inversion symmetry of the bulk Hamiltonian. On the other hand, the edge solitons are obtained by asymmetrically cutting the bulk solitons into two pieces, such that they individually have nonzero $S_z$ and break the sublattice inversion symmetry. This behaviour is strongly reminiscent of the linear limit, in which topological edge states are created when the edge cuts a dimer in half.

Whether a dimer is cut by the edge is determined by the Zak phase $\gamma$, which is quantized by inversion symmetry. When $\gamma = \pi$ (nontrivial phase) the Wannier centres lie at the cell boundary, corresponding to dimers cut by the edge. The Zak phase can be generalized to the nonlinear case by computing the nonlinear Berry phase of the delocalized modes comprising the bulk band structure~\cite{BerryPhaseNL2010}. For the nonlinear chain, provided the total dimer intensity  $I \equiv |\psi_1|^2 + |\psi_2|^2 $ is lower than the critical value $ I < 2 \eps / g$, the bulk band structure features two dispersive pass-bands,
$ \omega^{N\!L}_{\pm} (k_y) = \omega_{\pm}^{L}(k_y) - gI/2$, where linear dispersion $\omega_{\pm}^{L}(k_y)$ undergoes a uniform shift towards negative frequencies. The corresponding eigenfunctions ${\bf u}_{\pm} = \sqrt{\frac{I}{2}} \left( e^{-i \phi ( k_y ) }, \pm 1  \right)$ do not change compared to the linear case, so that $\gamma$ retains its same quantized values, i.e. $ \gamma_{\pm}=i \int_{-\pi}^{\pi} dk_y \langle \mathbf{u}_{\pm} | \frac{\partial \mathbf{u}_{\pm}}{\partial k_y} \rangle = \pi $ in the nontrivial case. While a bifurcation occurs above the critical intensity $I > 2 \eps / g$, creating an additional flat bulk band at frequency $\omega (|k_y|<\sqrt{(Ig/2)^2 - \eps^2}) = - g I$, this transition does not appear to affect the bulk solitons or the band gap, since it lies outside their frequency range of $[-\eps,\eps]$. Moreover, though localized modes are characterized by their dispersion $\mathcal{P}(\omega)$, which can be related to the local intensity at the edge as $\omega = - g I(y=0)/2$, this cannot be directly compared with the Bloch wave intensity $I$. 

These results suggest that the stationary bulk solitons are more relevant than the bulk nonlinear Bloch waves to the properties of edge states. In particular, edge states emerge at a symmetry-breaking bifurcation of the bulk solitons, which lose their stability. This picture is fully consistent with recent studies of topological laser arrays~\cite{Malzard2018,Cancellieri2019}, which proposed nonlinear topological transitions protected by particle-hole symmetry. By contrast, here the relevant symmetry is inversion symmetry~\cite{Grusdt2013}.

\begin{figure} [t!]
\centerline{\includegraphics*[width=8.6cm]{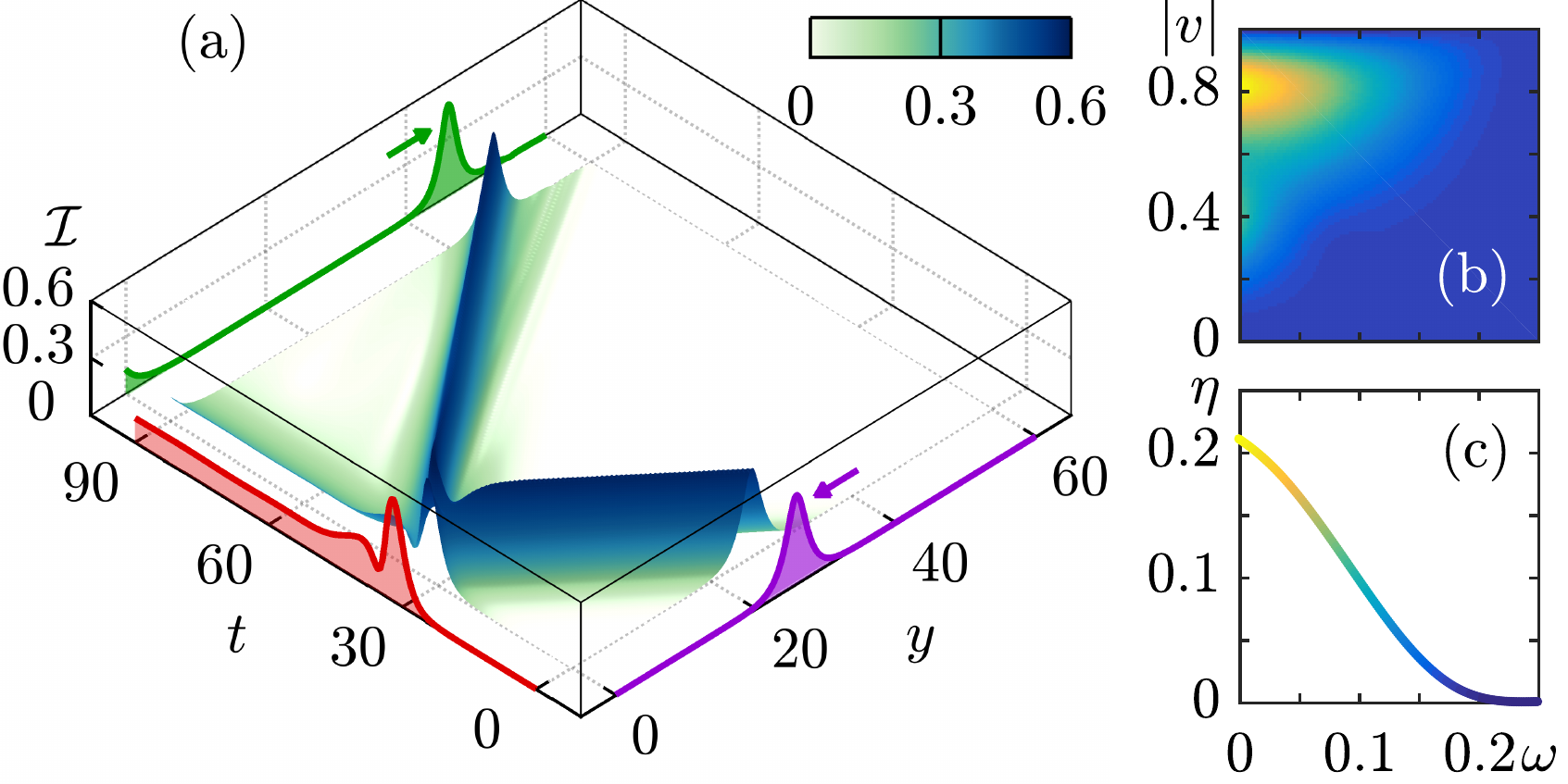}} 
\caption{
(a) Spatiotemporal dynamics in a finite dimer chain with a nontrivial 
edge $y=0$. Final and initial snapshots of the intensity distribution are shaded in green and purple, respectively. Parameters $\kappa=1$, $g=1$, $\eps = 0.25$, $\omega = 0.075$, $v = - 0.75$.
(b) Conversion efficiency to the edge state from an impinging stable solitons $\eta \equiv {\mathcal P}^{(1)}/{\mathcal P}_s$ depending on the soliton frequency and velocity. (c) Frequency dependence of the efficiency at fixed $v = -0.75$.  
}
 \label{fig:fig3}
\end{figure}

Eq.~\eqref{eq:soliton_spinor} can additionally describe moving solitons that are capable of exciting edge states by reflecting off the topologically nontrivial edge, as illustrated in Fig.~\ref{fig:fig3}. 
Upon reflection, the conservation of power and energy defines conservation laws for incident and reflected solitons and the excited edge state. The conversion efficiency tends to grow with decreasing frequency of the pump soliton, i.e. as the bifurcation point is approached, and for a given soliton frequency there is an optimal velocity maximizing the energy transfer. This effect holds beyond the continuum limit in finite discrete lattices, which is illustrated further in the Supplementary Material. 

The model~\eqref{eq:DiracEq1} can also be 
implemented in nonlinear photonic graphene with staggered sublattice potential.  
Near the Dirac points of the bandgap structure, the low-energy Hamiltonian assumes the form
\begin{equation} \label{eq:Hamilmass}
\hat{H}_{K_{\pm}} =  \pm \delta k_x \hat{\sigma}_x + \delta k_y \hat{\sigma}_y + M \hat{\sigma}_z\:,
\end{equation}
where $M$ is the detuning of the eigenfrequencies (mass) of two resonators in the bipartite unit cell of graphene.  Neglecting inter-valley scattering, we restrict our consideration to single valley $K_{+}$. 
The {\it valley-Hall domain wall} is created 
between two insulators characterized by distinct values of the mass, $M(y>0) = M_0$ and $M(y<0) = -M_0$. 
Such PT-symmetric interface with a real-valued mass governs the relation between the components of the edge state wavefunction that can be interpreted as an effective boundary condition~\cite{Ni2018}. For propagating waves $ \propto e^{-i\omega t + ikx}$ bound to the interface $y=0$, the relation $\psi_1(0) = \mp \psi_2(0)$ holds for $ M_0 > 0$ and $M_0 < 0$, respectively. Near the Dirac points, the valley edge states have the linear dispersion $ \omega = \mp k$ traversing the gap between the hyperbolic  cones of bulk states $\omega^2 = k^2 + M_0^2 $. Introducing nonlinearity, with the use of soliton solution~\eqref{eq:soliton_spinor}, we derive the nonlinear dispersion for the edge states
\begin{equation} 
\omega =  \Delta \omega_{N\!L} \mp k\:,
\end{equation}
where the Dirac cross undergoes a shift $\Delta \omega_{N\!L} = - \dfrac{g}{2} \varrho_s \left(\alpha_s = \pm \dfrac{\pi}{4} \right)$ in the band gap $[-M_0, M_0]$.
The valley Chern number formally stays the same as its linear counterpart until the upper band of nonlinear Bloch waves 
forms a self-crossing loop at above-threshold bulk intensities $I \ge 2 M_0$~\cite{Bomantara2017}. Thereby, nonlinearity grants control over the frequency and transverse structure of the edge state as defined by $\varrho_s(\alpha_s)$. We stress that our solution describes the localization transverse to the edge assuming plane wave-like profiles with fixed $k$ parallel to the edge. For finite wavepackets, higher order terms in $k$ will induce diffraction along the edge, with weak nonlinearity inducing edge solitons~\cite{Ablowitz2014}.

\begin{figure} [t!]
\centerline{\includegraphics*[width=8.6cm]{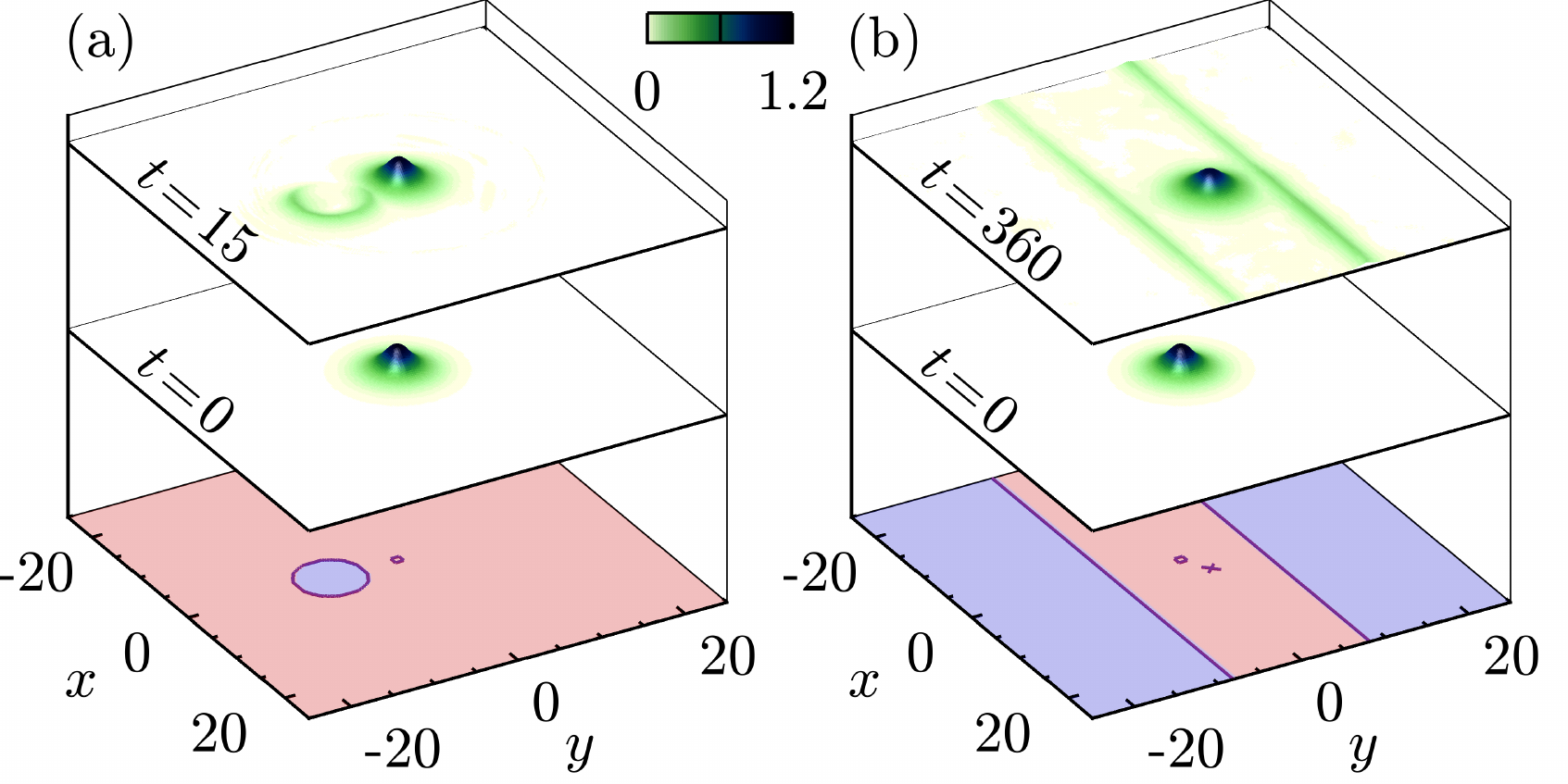}}  
\caption{
Solitons excite edge states at the domain walls created by mass inversion in a two-dimensional Dirac model. Bottom slices show mass maps: (a) a circular cavity, (b) two straight interfaces, with domains of positive (reddish) 
and negative (bluesh) 
masses. Parameters $M_0=1$, $g=1$, $\omega$ = 0.7. 
Middle and top slices: intensity distribution  ${\left(|\psi_1|^2 + |\psi_2|^2\right)^{\frac{1}{3}}}$ at the initial and given moment. Locations of the intensity maxima are marked with a dot ($t=0$) and cross ($t=360$).  
}
 \label{fig:fig4}
\end{figure}

In contrast to the 1D case, 2D bulk solitons at $M \ne 0$ cannot be obtained analytically. Nevertheless, by analogy with the above results we can consider excitation of valley edge sites by scattering of bulk solitons. Stationary bulk semi-vortex solitons with harmonic time dependence and radial symmetry, whose spinor in the polar coordinate system $(r,\varphi)$ is given by $[\Psi_{1},\Psi_{2}] = [u(r), iv(r)e^{i\varphi}]e^{-i\omega t }$ at $M_0>0$ and $[\Psi_{1},\Psi_{2}] = [-v(r)e^{-i\varphi}, iu(r)]e^{-i\omega t }$ at $M_0<0$, are numerically calculated by the shooting technique after Chebyshev discretization on the radial coordinate  is performed~\cite{Maravero2016,RingSoliton2018}.
At $M_0 = 1$, solitons are found to be stable at $\omega > 0.388$ in agreement with Refs.~\cite{Maraver2017b,Sakaguchi2018}.
Time dynamics is modeled 
with a custom numerical code employing a split-step scheme 
and the Fourier spectral method accomplished by means of the fast Fourier
transform. Periodic boundary conditions are applied to the 
rectangular calculation domain with an equispaced grid. In Fig.~\ref{fig:fig4}(a), a topological cavity is created by mass inversion in a circular domain. 
A clock-wise pulse of edge waves is seen to be excited at a closed contour of the cavity by a bulk soliton. The soliton excites persistent topological current and goes away 
tilted in the opposite direction. 
In Fig.~\ref{fig:fig4}(b), a soliton initially launched 
in $y$ direction is trapped in a waveguide made of two parallel topological interfaces supporting counter-propagating valley edge states. The soliton moves in a zigzag between the waveguide boundaries. Upon each reflection from the 
the walls, 
it emits a pulse of edge waves. 

Given the potential to explore the interplay between nonlinear and topological
effects in the photonic settings, where the infinite-contrast tight-binding approximation is often less applicable, the perspective of continuum nonlinear edge models and bifurcation analysis offer valuable insights. 
Our method can be  applied to other classes of topological lattices, with the example of the nonlinear distorted Kagome lattice given in Supplemental Material.

In summary, we have studied topological localized states in the nonlinear Dirac model and demonstrated close connections between edge states and self-trapped nonlinear modes in topological band gaps. Both can be inferred from phase portraits of the same nonlinear mapping. The bulk solitons 
resemble the Wannier functions in the linear limit, with nonlinear edge states emerging precisely when the bulk solitons are cut by the edge. This bifurcation of edge states is accompanied by the bulk soliton destabilizing. We furthermore demonstrated numerically that mutual transformations between edge and bulk states, forbidden in linear limit, can occur in the nonlinear regime in one- and two-dimensional systems. Our findings could have important implications for further developments of nonlinear topological systems not being limited to photonics but spanning through the fields of metamaterials to the topological physics of cold atoms in optical lattices.

\begin{acknowledgements}
This work was supported by the Australian Research Council, the Strategic Fund of the Australian National University, the Russian Foundation for Basic Research (Grants 18-02-00381, 19-52-12053) and the Institute for Basic Science in Korea (IBS-R024-Y1). 
\end{acknowledgements}


\begin{thebibliography}{33}
\newcommand{\enquote}[1]{``#1''}

\bibitem{Lu2016}
L.~Lu, J.~D. Joannopoulos, and M.~Solja{\v{c}}i{\'{c}}, \enquote{Topological
  states in photonic systems,} Nature Physics \textbf{12}, 626--629 (2016).

\bibitem{Ozawa2019}
T.~Ozawa, H.~M. Price, A.~Amo, N.~Goldman, M.~Hafezi, L.~Lu, M.~C. Rechtsman,
  D.~Schuster, J.~Simon, O.~Zilberberg, and I.~Carusotto, \enquote{Topological
  photonics,} Rev. Mod. Phys. \textbf{91}, 015006 (2019).

\bibitem{Zhou2017}
X.~Zhou, Y.~Wang, D.~Leykam, and Y.~D. Chong, \enquote{Optical isolation with
  nonlinear topological photonics,} New J. Phys. \textbf{19}, 095002 (2017).

\bibitem{Kartashov2017}
Y.~V. Kartashov and D.~V. Skryabin, \enquote{Bistable topological insulator
  with exciton-polaritons,} Phys. Rev. Lett. \textbf{119}, 253904 (2017).

\bibitem{Segev2018b}
M.~A. Bandres, S.~Wittek, G.~Harari, M.~Parto, J.~Ren, M.~Segev, D.~N.
  Christodoulides, and M.~Khajavikhan, \enquote{Topological insulator laser:
  Experiments,} Science \textbf{359}, eaar4005 (2018).

\bibitem{Chen2018}
W.~Chen, D.~Leykam, Y.~Chong, and L.~Yang, \enquote{Nonreciprocity in synthetic
  photonic materials with nonlinearity,} {MRS} Bulletin \textbf{43}, 443--451
  (2018).

\bibitem{Mittal2018}
S.~Mittal, E.~A. Goldschmidt, and M.~Hafezi, \enquote{A topological source of
  quantum light,} Nature \textbf{561}, 502 (2018).

\bibitem{Kruk2018}
S.~Kruk, A.~Poddubny, D.~Smirnova, L.~Wang, A.~Slobozhanyuk, A.~Shorokhov,
  I.~Kravchenko, B.~Luther-Davies, and Y.~Kivshar, \enquote{Nonlinear light
  generation in topological nanostructures,} Nature Nanotechnology \textbf{14},
  126--130 (2018).

\bibitem{Leykam2018}
D.~Leykam, S.~Mittal, M.~Hafezi, and Y.~D. Chong, \enquote{Reconfigurable
  topological phases in next-nearest-neighbor coupled resonator lattices,}
  Phys. Rev. Lett. \textbf{121}, 023901 (2018).

\bibitem{Wang2019}
S.~Wang, L.-J. Lang, L.~C. H., B.~Zhang, and Y.~D. Chong,
  \enquote{Topologically enhanced harmonic generation in a nonlinear
  transmission line metamaterial,} Nature Commun. \textbf{10}, 1102 (2019).

\bibitem{Bardyn2016}
C.-E. Bardyn, T.~Karzig, G.~Refael, and T.~C.~H. Liew, \enquote{Chiral
  bogoliubov excitations in nonlinear bosonic systems,} Phys. Rev. B
  \textbf{93}, 020502 (2016).

\bibitem{Bleu2016}
O.~Bleu, D.~D. Solnyshkov, and G.~Malpuech, \enquote{Interacting quantum fluid
  in a polariton chern insulator,} Phys. Rev. B \textbf{93}, 085438 (2016).

\bibitem{Peano2016}
V.~Peano, M.~Houde, F.~Marquardt, and A.~A. Clerk, \enquote{Topological quantum
  fluctuations and traveling wave amplifiers,} Phys. Rev. X \textbf{6}, 041026
  (2016).

\bibitem{Segev2013}
Y.~Lumer, Y.~Plotnik, M.~C. Rechtsman, and M.~Segev, \enquote{Self-localized
  states in photonic topological insulators,} Phys. Rev. Lett. \textbf{111},
  243905 (2013).

\bibitem{Leykam2016}
D.~Leykam and Y.~D. Chong, \enquote{Edge solitons in nonlinear-photonic
  topological insulators,} Phys. Rev. Lett. \textbf{117}, 143901 (2016).

\bibitem{Solnyshkov2017}
D.~Solnyshkov, O.~Bleu, B.~Teklu, and G.~Malpuech, \enquote{Chirality of
  topological gap solitons in bosonic dimer chains,} Phys. Rev. Lett.
  \textbf{118}, 023901 (2017).

\bibitem{Gerasimenko2016}
Y.~Gerasimenko, B.~Tarasinski, and C.~W.~J. Beenakker,
  \enquote{Attractor-repeller pair of topological zero modes in a nonlinear
  quantum walk,} Phys. Rev. A \textbf{93}, 022329 (2016).

\bibitem{Hadad2016}
Y.~Hadad, A.~B. Khanikaev, and A.~Al\`u, \enquote{Self-induced topological
  transitions and edge states supported by nonlinear staggered potentials,}
  Phys. Rev. B \textbf{93}, 155112 (2016).

\bibitem{Hadad2017}
Y.~Hadad, V.~Vitelli, and A.~Alu, \enquote{Solitons and propagating domain
  walls in topological resonator arrays,} ACS Photonics \textbf{4}, 1974--1979
  (2017).

\bibitem{Hadad2018}
Y.~Hadad, J.~C. Soric, A.~B. Khanikaev, and A.~Al{\`{u}}, \enquote{Self-induced
  topological protection in nonlinear circuit arrays,} Nature Electronics
  \textbf{1}, 178--182 (2018).

\bibitem{RingSoliton2018}
A.~N. Poddubny and D.~A. Smirnova, \enquote{Ring {D}irac solitons in nonlinear
  topological systems,} Phys. Rev. A \textbf{98}, 013827 (2018).

\bibitem{Bisianov2018}
A.~Bisianov, M.~Kremer, A.~Szameit, and U.~Peschel, \enquote{Experimental
  observation of the coupling of a nonlinear wave to a topological edge state,}
  in \enquote{Conference on Lasers and Electro-Optics,}  (Optical Society of
  America, 2018), p. FM1E.1.

\bibitem{Dobrykh2018}
D.~A. Dobrykh, A.~V. Yulin, A.~P. Slobozhanyuk, A.~N. Poddubny, and Y.~S.
  Kivshar, \enquote{Nonlinear control of electromagnetic topological edge
  states,} Phys. Rev. Lett. \textbf{121}, 163901 (2018).

\bibitem{BerryPhaseNL2010}
J.~Liu and L.~B. Fu, \enquote{Berry phase in nonlinear systems,} Phys. Rev. A
  \textbf{81}, 052112 (2010).

\bibitem{Malzard2018}
S.~Malzard, E.~Cancellieri, and H.~Schomerus, \enquote{Topological dynamics and
  excitations in lasers and condensates with saturable gain or loss,} Opt.
  Express \textbf{26}, 22506--22518 (2018).

\bibitem{Cancellieri2019}
E.~Cancellieri and H.~Schomerus, \enquote{$\mathcal{PC}$-symmetry-protected
  edge states in interacting driven-dissipative bosonic systems,} Phys. Rev. A
  \textbf{99}, 033801 (2019).

\bibitem{Grusdt2013}
F.~Grusdt, M.~H\"oning, and M.~Fleischhauer, \enquote{Topological edge states
  in the one-dimensional superlattice bose-hubbard model,} Phys. Rev. Lett.
  \textbf{110}, 260405 (2013).

\bibitem{Ni2018}
X.~Ni, D.~Smirnova, A.~Poddubny, D.~Leykam, Y.~Chong, and A.~B. Khanikaev,
  \enquote{$\mathcal{PT}$ phase transitions of edge states at $\mathcal{PT}$
  symmetric interfaces in non-hermitian topological insulators,} Phys. Rev. B
  \textbf{98}, 165129 (2018).

\bibitem{Bomantara2017}
R.~W. Bomantara, W.~Zhao, L.~Zhou, and J.~Gong, \enquote{Nonlinear {D}irac
  cones,} Phys. Rev. B \textbf{96}, 121406 (2017).

\bibitem{Ablowitz2014}
M.~J. Ablowitz, C.~W. Curtis, and Y.-P. Ma, \enquote{Linear and nonlinear
  traveling edge waves in optical honeycomb lattices,} Phys. Rev. A
  \textbf{90}, 023813 (2014).

\bibitem{Maravero2016}
J.~Cuevas\char21{}Maraver, P.~G. Kevrekidis, A.~Saxena, A.~Comech, and R.~Lan,
  \enquote{Stability of solitary waves and vortices in a 2{D} nonlinear {D}irac
  model,} Phys. Rev. Lett. \textbf{116}, 214101 (2016).

\bibitem{Maraver2017b}
J.~Cuevas-Maraver, P.~G. Kevrekidis, A.~B. Aceves, and A.~Saxena,
  \enquote{Solitary waves in a two-dimensional nonlinear {D}irac equation: from
  discrete to continuum,} J. Phys. A \textbf{50}, 495207 (2017).

\bibitem{Sakaguchi2018}
H.~Sakaguchi and B.~A. Malomed, \enquote{One- and two-dimensional gap solitons
  in spin-orbit-coupled systems with {Z}eeman splitting,} Phys. Rev. A
  \textbf{97}, 013607 (2018).
\end{thebibliography}

\end{document}